\documentclass[a4paper]{article} 
\usepackage{graphicx}
\usepackage[top=3.0cm,bottom=3.0cm,left=3.0cm,right=2.0cm]{geometry}
\usepackage{amssymb,amsmath}
\usepackage{textcomp}
\usepackage{subfigure}
\usepackage{cite}
\usepackage{braket}
\usepackage{cleveref}
\usepackage{slashed}
\usepackage{stackrel}
\usepackage[0]{editing}
\usepackage{kantlipsum}
\usepackage{sidecap}

\usepackage[shortcuts,acronym,nonumberlist,nomain]{glossaries}

\providecommand{\keywords}[1]{\textbf{{Keywords:}} #1}

\title{Transport Equation for Small Systems and Nonadditive Entropy}
\author{Eugenio Meg\'{\i}as$^{1}$, Jose A. S. Lima$^{2}$ and Airton Deppman$^{3}$}

        \date{
        \begin{small}
    $^{1}$ Departamento de F\'{\i}sica At\'omica, Molecular y Nuclear and Instituto Carlos I de F\'{\i}sica Te\'orica y Computacional, Universidad de Granada, Avenida de Fuente Nueva s/n, 18071 Granada, Spain; emegias@ugr.es \\
$^{2}$ Instituto de Astronomia e Geof\'{i}sica da Universidade de S\~ao Paulo, Rua do Mat\~ao 1371-Butant\~a, S\~ao Paulo-SP, CEP 05580-090, Brazil;  jas.lima@iag.usp.br \\
$^{3}$ Instituto de F\'{\i}sica, Universidade de S\~ao Paulo, Rua do Mat\~ao 1371-Butant\~a, S\~ao Paulo-SP, CEP 05580-090, Brazil; deppman@usp.br\\
       \end{small}
       \vspace{0.5cm}
       }

\begin{document}

\maketitle

\bigskipamount=1cm

\abstract{The nonadditive entropy introduced by Tsallis in 1988 has been used in different fields and generalizes the Boltzmann entropy extending the possibilities of application of the statistical methods developed in the context of Mechanics. Here we investigate one of the last points of the theory that still are under discussion: the source term of the nonextensive transport equation. Based on a simple system, we show that the nonadditivity is a direct consequence of the phase space topology, and derive the source term that leads to the nonextensive transport equation.}

\vspace{0.5cm}

\section{Introduction}

The concept of entropy has been continuously developed since its introduction by Clausius. Boltzmann gave it a precise meaning in Statistical Mechanics, and Shannon extended the reach of the concept to Information Sciences. The entropy defined by Boltzmann was generalized by Tsallis in a seminal paper~\cite{Tsallis}, prompting many works that deepened our understanding of the physical and statistical meaning of this important quantity. One of the most distinct features of the generalized entropy is that it breaks with the additivity of the Boltzmann entropy. The new entropy is defined as
\begin{equation}
 S_q(p_i)=\frac{k}{1-q}\left( \sum_i p_i^q -1 \right) \,,
\end{equation}
where $p_i$ is the probability for the event $i$, $k$ is the Boltzmann constant, and $q$ is a parameter called the entropic index. The parameter $q$ is a measure of the nonadditivity of the system. For $q \rightarrow 1$, the usual Boltzmann entropy is recovered. The nonadditivity becomes evident when we consider the composition of two nonextensive systems, $A$ and $B$, such that their entropies are $S(A)$ and $S(B)$, respectively. The entropy of the system resulting from the combination of $A$ and $B$ is
\begin{equation}
 S_q(AB)=S_q(A)+S_q(B)+\frac{(1-q)}{k} S_q(A)S_q(B)\,.
\end{equation}

Many studies followed the paper of Tsallis, evaluating the different methods one can adopt to generalize the Boltzmann entropy. Superstatistics is a very general formalism to obtain different nonextensive statistics~\cite{Beck-Cohen-Superstatistics}. The axiomatic approach used in~\cite{HanelThurner} clarifies the mathematical aspects of the generalized statistics, as well as the group-theoretical approach~\cite{Tempesta}. A comprehensive account of the mathematical aspects of the generalization of entropy can be found in~\cite{Umarov-Tsallis}

A few physical mechanisms that could result in the nonextensive statistics have been proposed. Temperature fluctuations are one of the most known mechanisms that gives the origin to $q$-exponential distributions typical of the Tsallis statistics~\cite{Wilk:2012zn}. The small system effects have been studied in many works~\cite{Biro:2020kve}. The $q$-exponential Tsallis distribution can be uniquely determined from two simple conditions~\cite{Silva98}: (i) isotropy
of the velocity space and (ii) a suitable generalization of the Maxwell factorizability condition, or equivalently, the assumption that $f(v) \neq f(v_x)f(v_y)f(v_z)$. Another approach uses a particular form of the Fokker--Planck equation~\cite{Lisa-Borland}. In this connection, the classical work of Chandrasekhar~\cite{Chandra} on dynamic friction in the gravitational context has also been extended based on Tsallis' distribution~\cite{Silva-Lima}. Finally, fractal structures can also explain the origin of the Tsallis distribution in a general class of systems~\cite{Deppman-PRD-2016}.

There is a large number of applications of the Tsallis statistics in Physics and in other domains, as can be seen in~\cite{Tsallis-book} and the references therein. In some cases, the entropic parameter, $q$, which is a measure of the nonadditive of the entropy, can be theoretically calculated. This is the case of the spin glass system investigated in~\cite{CarusoTsallis}, the velocity distribution of cold atoms~\cite{Lutz}, anomalous diffusion~\cite{TsallisBukman}, black hole thermodynamics~\cite{Mejrhit:2020dpo}, and Yang--Mills field theory~\cite{DMM-PRD-2020}. In the latter, for QCD in particular, the entropic parameter can be calculated by a simple formula that involves only the fundamental parameters of the theory, the number of colors, $n_C=3$, and the number of flavors, $n_f=6$. The theoretical value predicted in~\cite{DMM-PRD-2020} gives $q=8/7$, which results in good agreement with the analyses of the experimental data on the particle production in high-energy collisions~\cite{Cleymans-Worku2,Cleymans-Parvan,Wilk-Wlodarczyk1,Li-Wang-Liu,T-Biro2,Shen}. Furthermore, in~\cite{Walton-Rafelski}, the Tsallis parameter was obtained by fitting the value of $q$ to adjust the Fokker--Planck dynamics equation of a quark moving in the QGP medium to fit the calculations obtained through a phenomenological QCD calculation with $n_f=3$, obtaining $q \sim 1.1$. Using this value for the number of flavors in the formula found in~\cite{DMM-PRD-2020} gives a result in agreement with the phenomenological approach. These results are enough to dismiss any claim that the parameter $q$ has no physical meaning.

Despite the successful application of the Tsallis statistics in many areas,~\cite{Kapusta} summarized a few criticisms, all of them fully addressed in a reply by C. Tsallis~\cite{Tsallis-replyKapusta}. One of those criticisms refers to the transport equation associated with the nonextensive statistics presented in~\cite{Lavagno}, which follows the one proposed in~\cite{Lima-Plastino-PRL}. In this work, we investigate the fundamental relation between the phase space topology and the nonadditive entropy introduced by Tsallis. We investigate the simplest problem of Statistical Mechanics, namely a system composed of a finite number of point-like particles, to show that, contrary to the usual expectation, this system follows the Tsallis statistics instead of the Boltzmann one. We show that the entropy cannot be additive in general and then investigate the consequences of the nonadditive distributions in the transport equation.

In the following, we address two aspects of the Tsallis statistics. We show that nonadditivity is present in the simplest possible statistical system. Then, we investigate how the nonadditivity modifies the source term of the transport equation. It is important to mention that the nonextensive transport equation has been studied in other works~\cite{Lima-Plastino-PRL,Lavagno}, but no explanation was given to the source term used. Here, we show how the topological properties of the phase space lead to a source term that reduces to that proposed in~\cite{Lavagno} under appropriate conditions. Our main objective is to show that the topology of the phase space leads to the nonextensive distributions in the form of $q$-exponential functions and that the $q$-algebra~\cite{BORGES-qCalculus} must be used. Then, we discuss the implications of these results in the source term of the transport equation. The association of the Symplectic Geometry with Group Theory is a rich field of research, and it is particularly useful in Quantum Theory, where the Hopf algebra (also known as quantum algebra or $q$-algebra) has been used in the context of the Bogoliubov transformations~\cite{Bogoliubov,Majid,Segovia1}. Here, we deal with the $q$-calculus associated with the Tsallis entropy.

In Section~\ref{sec2}, we address the statistical aspects of small systems by investigating the number of possible configurations for an ideal gas with a fixed and finite number of particles. In Section~\ref{sec3}, we show that the composition law of two small systems is governed by the $q$-calculus, and in Section~\ref{sec4}, we describe how such a composition law affects the collision term of the transport equation. In Section~\ref{sec5}, we present our conclusions.

\section{Statistical Mechanics of Small Systems}
\label{sec2}

The effects of the limited size of the system that give origin to the Tsallis statistics are well known~\cite{Biro:2020kve,Lima-Deppman-2020}. Here, we follow a procedure that is similar, but more general than that used in~\cite{Lima-Deppman-2020}, where the emergence of the $q$-exponential distribution was shown in an explicit way.
For the sake of simplicity, we investigate a non-relativistic system. The case of a relativistic system was treated in~\cite{DMM-PRD-2020}.

Consider a system of $N$ point-like particles with mass $m$ in a container with volume $V$, free of any external force. At equilibrium, the number of configurations for the system with total energy $E$ is
\begin{equation}
M =V \int d^{3N}p \, \delta(E_p-E) \,,
\end{equation}
where
\begin{equation}
 E_p=\frac{1}{2m}\sum_{i=1}^{3N} p_i^2= \frac{p^2}{2m} \,,
\end{equation}
and we define $p^2=\sum_{i=1}^{3N} p_i^2$. The usual procedure to calculate the integral is to use the transformation to spherical coordinates and integrate over the angles using the isotropic symmetry of the problem, i.e., 
\begin{equation}
\int_{\Omega_p} d^{3N}p = dp \, p^{3N-1} \int d\Omega_p = dp \, p^{3N-1} \, S(3N) \,, \label{transformation}
\end{equation}
where $\int_{\Omega_p}$ stands for the integral in the angles of the momentum, and
\begin{equation}
 S(n)=\frac{2 \pi^{n/2}}{\Gamma(n/2)} \,.
\end{equation}
The integral in Eq. (3) can be easily performed with the use of the delta-function, resulting in
\begin{equation} 
 M = V S(3N) m (2mE)^{3N/2-1} \,. \label{dMdE}
\end{equation}

If $A$ and $B$ are two partitions of the system with $N_A$ and $N_B$ particles and energies $E_A$ and $E_B$, respectively, such that $N_A+N_B=N$ and $E_A+E_B=E$, the number of configurations in which these partitions can appear is
\begin{equation}
 \frac{dM}{dE_A} = V \times \left[ S(3N_A) m (2mE_A)^{3N_A/2-1} \right] \times \int d^{3N_B}p_B \, \delta(E_{A} + E_{B} -E)\,, \label{eq:dMdEA}
\end{equation}
where
\begin{equation}
 E_{A} = \sum_{i=1}^{3N_A} \frac{p_{A i}^2}{2m} = \frac{p_A^2}{2m} \,, \qquad \textrm{and} \qquad E_{B} = \sum_{i=1}^{3N_B} \frac{p_{B i}^2}{2m} = \frac{p_B^2}{2m} \,.
\end{equation}

In obtaining Equation~(\ref{eq:dMdEA}), we integrated out just in the angles of the momentum $\vec{p}_A$, i.e., $\int d\Omega_{p_A} = S(3N_A)$, but not in its modulus. Aside from the number of particles and the energy, the only constraint to the partitions is that the total momentum of each partition is null in the frame of reference of the system. Using the transformation in~(\ref{transformation}), we have
\begin{equation} 
 \frac{dM}{dE_A}=V \times [S(3N_A) m (2mE_A)^{3N_A/2-1}] \times [S(3N_B) m (2mE_B)^{3N_B/2-1}] \,,
\end{equation}
where $E_B = E - E_A$. Clearly, $dM/dE_A$ represents only a fraction of the total configurations $M$. If $\rho(N_A,E_A)$ is the probability density to have the specific partition of the system, then $\rho(N_A,E_A) = (dM/dE_A)/M$. Therefore, using the equations obtained above, we have
\begin{equation}
 \rho(N_A,E_A)dE_A = \frac{\Gamma(3N/2)}{\Gamma(3N_A/2)\Gamma(3N_B/2)} \frac{dE_A}{E} \left(\frac{E_A}{E}\right)^{3 N_A/2 - 1} \left[1-\frac{E_A}{E}\right]^{3 N_B/2 - 1} \,. \label{probdensity}
\end{equation}

The term depending on the number of particles represents the combinatorial number of possible configurations for the partitions with $N_A$ and $N_B$ particles. The terms depending on the energies $E$ and $E_A$ are due to the topology of the phase space available to the system.

Observe that the probability density depends only on $N$, $N_A$, and $N_B$ and on the ratio $\chi=E_A/E$. This is a key aspect of the calculations that allows its extension to fractal systems~\cite{DMM-Federico-2018}. The $q$-exponential behavior can be already observed in the equation above, but we can make it more explicit by defining $(1-q_A)^{-1}=3N_B/2-1$ and $kT_A = (1-q_A) E$.
Then, the probability density can be written as
\begin{equation}
 \rho(N_A,E_A) dE_A = \frac{\Gamma(3N/2)}{\Gamma(3N_A/2)\Gamma(3N_B/2)} \frac{dE_A}{E} \left(\frac{E_A}{E}\right)^{3N_A/2-1} \left[1-(1-q_A)\frac{E_A}{kT_A}\right]^{\frac{1}{1-q_A}} \,.
\end{equation}

The $q$-exponential term appears because of the topological structure of the phase space for a system with a limited number of components. The use of the $q$-exponential distribution in high-energy data analysis and the physical meaning of the temperature in that distribution was described in~\cite{Biro:2020kve}.

We use the relation~(\ref{transformation}) backward to obtain
\begin{equation} \rho(N_A,E_A) dE_A= \frac{\Gamma(3N/2)}{\pi^{3N_A/2}\Gamma(3N_B/2)} \int_{\Omega_{p_A}} d^{3N_A}(p_A/p) \left[1-(1-q_A)\frac{E_A}{kT_A}\right]^{\frac{1}{1-q_A}} \,, \label{nonextensvedensity}
\end{equation}
where $p = \sqrt{2mE}$. Here, it is evident that the $q$-exponential term plays a role identical to the exponential Boltzmann factor in the standard statistics for physical systems. For $q\rightarrow 1$, which implies $N \rightarrow \infty$, the Boltzmann statistics is recovered, as expected in the Tsallis generalization. Thus, the Tsallis statistics is the correct framework to deal with the statistics of finite physical systems. For $N_A=1$, Equation~(\ref{nonextensvedensity}) reduces to the result found in~\cite{Lima-Deppman-2020}. This result has been interpreted as an indication that the Tsallis statistics is not only more general, but also more fundamental than the Boltzmann statistics. The limit $q \sim 1$ is obtained very fast, for the present case of a system formed by point-like particles, as shown by the relation between $q$ and $N$. This happens because the number of degrees of freedom (ndf) is proportional to the number of particles. For thermofractals, however, the ndf is independent of the number of components, which confers to the system a nonextensive behavior that is independent of the system size.

\section{Relation with the \boldmath$q$-Algebra}

\label{sec3}
The same procedure used to find the energy fluctuation of the partition $A$ can be used for the partition $B$. In this case, we would obtain
\begin{equation}
 \rho(N_B,E_B) dE_B = \frac{\Gamma[3N/2]}{\pi^{3N_B/2} \Gamma(3N_A/2)} \int_{\Omega_{p_B}} d^{3N_B}(p_B/p) \left[1-(1-q_B)\frac{E_B}{kT_B}\right]^{\frac{1}{1-q_B}} \,.
\end{equation}

Let us consider that the partitions $A$ and $B$ combined form a third partition $C$, with $N_C=N_A+N_B$ particles and energy $E_C=E_A+E_B$. The partition $C$ is initially in thermal contact with a system $O$ with $N_O$ particles and energy $E_O$ at temperature $T_C$. 
The energy fluctuation of the system C is
\begin{equation}
 \rho(N_C,E_C) dE_C = C(N_C) \int_{\Omega_{p_C}} d^{3N_C}\left( p_C/p \right) \left[1-(1-q_C)\frac{E_C}{kT_C}\right]^{\frac{1}{1-q_C}} \,, \label{finitesystemcorrelation}
\end{equation}
where $C(N_C)$ is a multiplication term that depends only on $N_C$, and \mbox{$(1-q_C)^{-1}=3N_O/2-1$.} The energies of the partitions $A$ and $B$ of the system $C$ also fluctuates, but we consider that the number of particles in each partition, $N_A$ and $N_B$, does not change. Let us say that at the instant $t=0$, the system $C$, which was initially in thermal contact with the system $O$, is isolated. At this moment, the system $C$ has energy $E_C$, while the partitions of $C$ have energies $E_A$ and $E_B$, respectively. By using
\begin{equation}
\rho(N_C,E_C) \simeq \int_0^{E_C} dE_A dE_B \, \rho(N_A,E_A) \rho(N_B,E_B) \delta(E_A + E_B - E_C) \,,
\end{equation}
%
an approximation that turns out to be better for $N_O \gg N_C$, one can see that the fluctuations of the three systems are not independent, but are related by
%
\begin{equation}
 \left[1-(1-q_C)\frac{E_C}{kT_C}\right]^{\frac{1}{1-q_C}} \sim \left[1-(1-q_A)\frac{E_A}{kT_A}\right]^{\frac{1}{1-q_A}} \times \left[1-(1-q_B)\frac{E_B}{kT_B}\right]^{\frac{1}{1-q_B}} \,.
\end{equation}

Since the partitions $A$ and $B$ are still in thermal contact with each other, their energies can fluctuate to $\bar E_A$ and $\bar E_B$, but they still have to satisfy the relation
\begin{equation}
 \left[1-(1-q_C)\frac{E_C}{kT_C}\right]^{\frac{1}{1-q_C}} \sim \left[1-(1-q_A)\frac{\bar E_A}{kT_A}\right]^{\frac{1}{1-q_A}} \times \left[1-(1-q_B)\frac{\bar E_B}{kT_B}\right]^{\frac{1}{1-q_B}} \,. \label{qproduct}
\end{equation}

Here, it is evident that the composition of the two subsystems, $A$ and $B$, is statistically described by a product law that is not the usual algebraic multiplication law. The correct algebra does deal with the composition of phase space statistics, the so-called $q$-algebra developed in the context of the Tsallis statistics~\cite{BORGES-qCalculus,Nivanen}. Denoting by $f(E)$ the $q$-exponential function and using the relation between $q_j$ and $T_j$, with $j=A,B,C$, we observe that
\begin{equation}
 \begin{cases}
  f_A^{1-q_A}= 1-E_A/E \\
  f_B^{1-q_B}= 1-E_B/E \\
  f_C^{1-q_C}= 1-E_C/E \,;
 \end{cases}
\end{equation}
therefore, $f_C^{1-q_C}=f_A^{1-q_A}+f_B^{1-q_B}-1$, where we used $E_C=E_A+E_B$. It follows that the composition law must be
\begin{equation}
 f_C=\left[f_A^{1-q_A}+f_B^{1-q_B}-1\right]^{\frac{1}{1-q_C}} \,, \label{compositionlaw}
\end{equation}
where $f_X = f(E_X)$ for $X = A, B, C$. An analysis of the transformation properties of the phase space, in the context of the thermofractals, and its relations with the $q$-algebra can be found in~\cite{Deppman-Physics-2021}. If we take $N\rightarrow \infty$, $N_A\rightarrow \infty$, and $N_B\rightarrow \infty$, the $q$-exponential distributions of $E_C$, $E_A$, and $_B$, respectively, turn into exponential functions, and Equation~(\ref{qproduct}) reduces to the ordinary product of exponential functions. We do not claim that the results obtained above represent an extension of the Tsallis statistics to systems with a variable entropic index. Here, $q(N)$ changes when the number of particles in the system changes. Our objective here is to show that the dynamical evolution of the small system considered has a source term that satisfies Equation~(\ref{compositionlaw}) and that such a source term derives from the topology of the phase space. 

The connections between the phase space topology and the Tsallis statistics are not restricted to the $q$-exponential distribution, but go as deep as the $q$-algebra associated with the nonadditivity of the entropy. The causes of these close relations can be understood in light of the present study. There are two constraints that impose a correlation between the systems $A$ and $B$: one is evidently the energy conservation, which appears in the delta function in the distribution calculation; the other is the momentum conservation, which appears in a less evident way when we assumed the isotropic momentum distributions to integrate into the angles, that is when we perform the transformation~(\ref{transformation}).

\section{Transport Equation for Small Systems}
\label{sec4}

Here, we investigate the transport equation in the partition $C$ described above after the system is isolated at $t=0$. In the following, we assume that the Hamiltonian for the particles in the system $C$ is given by
\begin{equation}
 H_C=\sum_{i=1}^{3N_C} \frac{p_i^2}{2m} +\sum_{i,j = 1}^{N_C} V_{i,j} + \sum_{i=1}^{N_C} U_i + \sum_{i \in C\,,\,j \notin C} V_{i,j} \,.
\end{equation}

The interaction term, $V_{i,j}$, represents the elastic contact interaction among the particles of the system; $U_i$ represents the elastic interaction of particles and the heavy container wall; the last term in the right-hand side corresponds to the interaction between the system $C$ and the rest of the $N$-particle system.

The Hamiltonian above is very general, but we introduce the following constraints: (i)~the interaction range is small compared with the system dimension and the distance among the particles; (ii) in an interval of time $\delta t$ sufficiently small, the number of interactions is large enough to consider that the total momentum of the systems $A$, $B$, and $C$ does not vary, that is
\begin{equation}
 \vec P_X=0 \,, \qquad \textrm{for} \qquad X=A, B, C \,.
\end{equation}

At the instant $t=0$, the interaction between the system $C$ and the rest of the system is switched off, so the last term in the Hamiltonian is null. The average potential energy is null, since we are considering contact interactions between particles and between particles and walls. The only effect of the interaction is to transfer momentum from particles in $A$ to particles in $B$ and vice versa or to reverse one component of the particle momentum in the interaction with the wall. Under such conditions,
\begin{equation}
 E_C=H=\sum_{i=1}^{3 N_C} \frac{p_i^2}{2m}\,.
\end{equation}

The energies $E_A$ and $E_B$ are given by
\begin{equation}
 E_X =\sum_{i = 1}^{3N_X} \frac{p_{X i}^2}{2m}\,, \qquad \textrm{where} \qquad X = A, B \,,
\end{equation}
and $\Delta E_A=-\Delta E_B$, since the system $C$ is now isolated.

At the instant $t=\delta t$, the probability density for the system $C$, $F_C$, is still that given by Equation~(\ref{finitesystemcorrelation}), but the densities for the systems $A$ and $B$ have changed to $\bar f_A$ and $\bar f_B$, respectively, since their energies changed to $\bar E_A=E_A+\Delta E_A$ and $\bar E_B=E_B-\Delta E_A$, respectively.

If $f(\vec{x},\vec{p},t)$ is the local density of the system $C$ in the phase space at the instant $t \le \delta t$, the transport equation is~\cite{GMKremer}
\begin{equation}
 \frac{df}{dt}=G(f,\bar f)\,,
\end{equation}
where $G(f,\bar f)$ is the source term. The right-hand side can be expanded in the form
\begin{equation}
 \frac{df}{dt}=\frac{\partial f}{\partial t}+\vec{\nabla}_x f \cdot \vec{v} + \vec{\nabla}_v f \cdot \vec{F} \,.
\end{equation}

The small system effects are restricted to the source term, as we describe now.

The source term describes how the interaction among the systems $A$ and $B$ modifies the probability density $f$ of each system. For the simple case of elastic scattering among point-like particles, it is given by
\begin{equation}
 \begin{split}
 G(f,\bar f)= & \int d^3p_A \,~ f_A \, \int d^3p_B \, ~f_B \, \sigma(AB\rightarrow \bar A \bar B) \int d^3 \bar p_A \, ~ \bar f_A \, \int d^3 \bar p_B \,~ \bar f_B - \\ &\int d^3p_A \,~ f_A \, \int d^3p_B \,~ f_B \, \sigma(\bar A \bar B \rightarrow AB) \int d^3 \, \bar p_A \bar f_A \, \int d^3 \bar p_B \, \bar f_B \,, \label{eq:Gff1}
 \end{split}
\end{equation}
where $p_i$ and $\bar p_i$ are the components of the momentum of the system $i=A,B$, before and after the collision, while $f_X \equiv f(p_X)$ and $\bar f_X \equiv f(\bar p_X)$ for $X = A, B$. The symbol $\sigma(X \rightarrow Y)$ represents the cross-section for the process $X \rightarrow Y$.
\begin{equation}
 \begin{split}
 G(f,\bar f)= \int d^3p_A \int d^3p_B \int d^3 \bar p_A \int d^3 \bar p_B \, & [\sigma(AB \rightarrow \bar A \bar B) \, h[f_A,f_B] -  \sigma(\bar A \bar B \rightarrow AB) \, h[\bar f_A,\bar f_B]]  \,, \label{eq:Gff2}
 \end{split}
\end{equation}
where $h[f_A,f_B]$ is the correlation function of the systems $A$ and $B$.

The composition of the probability densities must follow Equation~(\ref{compositionlaw});
thus,
\begin{equation}
 h[f,\bar f]=\left[f_A^{1-q_A}+f_B^{1-q_B}-1\right]^{\frac{1}{1-q_C}}\,.
 \label{eq:newcomposition}
\end{equation}

Thus, the source term for a small system is different from the one used in the standard statistics. Naturally, when $N_A$ and $N_B$ are sufficiently large, this expression reduces to the usual Boltzmann source term:
\begin{equation}
 h[f,\bar f]=f_Af_B \,.\label{eq:standard}
\end{equation}

One of the consequences of Equation~(\ref{eq:newcomposition}) is that, in the stationary state, the function $f_j$ is a $q$-exponential, while for the standard composition law given by Equation~(\ref{eq:standard}), this leads to the exponential function. 

For the special case when $N_A=N_B$, we have a relation that is very close to the one proposed in~\cite{Lima-Plastino-PRL,Lavagno}. The difference with respect to this latter result is because the total number of particles in the system is twice the number $N_A$, so $q \ne q_A$. For thermofractals, where the pdf is independent of the number of particles, the combination rule in Equation~(\ref{compositionlaw}) would reduce to that proposed in~\cite{Lima-Plastino-PRL,Lavagno}. The relations between the source term and the H-theorem were explored in~\cite{Lima-Plastino-PRL}.

\section{Conclusions}
\label{sec5}

In this paper, we addressed the problem of the collision term of the transport equation for small systems. We used the simplest possible system, an ideal gas with a finite number of point-like particles, to study how the topological characteristics of the phase space determine the statistical aspects of the possible configurations of the system.

The statistical analysis of the system configurations showed that the composition of two partitions of the system is not additive, as happens in the thermodynamical limit, but follows a composition rule that corresponds to the $q$-calculus algebra.

We used the result of the statistical analysis and the resulting composition rule to show that it imposes a new collision term for the transport equation for those systems. The form of the collision term is similar to those adopted in~\cite{Lima-Plastino-PRL,Lavagno}. The new source term leads, in the equilibrium, to the $q$-exponential distribution instead of the exponential distribution obtained in the standard collision term (see also~\cite{Silva98}).

The main result of the present work demonstrated that, contrary to what is generally assumed, the new form of the collision term is a result of the topological properties of the phase space and results from fundamental aspects of the physical systems. 

Some potential applications of this work include the study of dissipative transport coefficients in kinetic theory~\cite{Arnold:2002zm}, as well as black hole thermodynamics, where the entropy is proportional to the area of the event horizon, but still additive~\cite{Mejrhit:2019oyi,Mejrhit:2020dpo,Nojiri:2021czz}. It was claimed in some of these references that the entropic index $q$ could play an important role in the stability of black holes, in analogy with previous studies in AdS space by using standard Boltzmann statistics. This leads also to many potential applications of the nonextensive statistics in the context of the AdS/CFT correspondence, in particular to study the strongly coupled regime of gauge theories~\cite{Ammon:2015wua}. At this point, the theoretical groups methods can be of great help to extend the $q$-algebra associated with the Tsallis entropy to study more complex mathematical structures~\cite{Segovia2}.

\vspace{6pt}


\section*{Acknowledgments}
The authors are thankful to Constantino Tsallis for fruitful
discussions. A.D. is supported by the Project Instituto Nacional de
Ci\^encia e Tecnologia-F\'{\i}sica Nuclear Aplicada INCT-FNA)
Proc. No. 464898/2014-5, by the Conselho Nacional de Desenvolvimento
Cient\'{\i}fico e Tecnol\'ogico (CNPq-Brazil), Grant 304244/2018-0, by
Project INCT-FNA Proc. No. 464 898/2014-5, and by FAPESP, Brazil,
Grant 2016/17612-7. The work of E.M. is supported by the project
PID2020-114767GB-I00 financed by MCIN/AEI/10.13039/501100011033, by
the FEDER/Junta de Andaluc\'{\i}a-Consejer\'{\i}a de Econom\'{\i}a y
Conocimiento 2014-2020 Operational Program under Grant A-FQM178-
UGR18, by Junta de Andaluc\'{\i}a under Grant FQM-225, and by the
Consejer\'{\i}a de Conocimiento, Investigaci\'on y Universidad of the
Junta de Andaluc\'{\i}a and European Regional Development Fund (ERDF)
under Grant SOMM17/6105/UGR. The research of E.M. is also supported by
the Ram\'on y Cajal Program of the Spanish MCIN under Grant
RYC-2016-20678. J.A.S.L. is partially supported by CNPq
(310038/2019-7), CAPES (88881.068485/2014), and FAPESP (LLAMA Project
No. 11/51676-9).



\linespread{.9}
\bibliographystyle{ieeetr}

\end{document}